\pdfoutput=1
\documentclass[conference]{IEEEtran}
\usepackage{cite}
\usepackage{url}
\usepackage{amsmath,amssymb,amsfonts}
\usepackage{algorithmic}
\usepackage{graphicx}
\usepackage{textcomp}
\usepackage{xcolor}
\usepackage{multirow}
\usepackage{subcaption}
\usepackage{tikz}
\usepackage{mfirstuc}
\usepackage[normalem]{ulem}
\usepackage{soul}
\usepackage{float}
\usepackage{multicol}
\usepackage{placeins}
\usepackage{subcaption}
\usepackage{comment}
\usepackage{hyperref}

\usepackage{caption}

\usepackage{xcolor}

\begin{document}
\title { Predicting DC-Link Capacitor Current Ripple in AC-DC Rectifier Circuits Using Fine-Tuned Large Language Models}

\author{Mohamed Zeid,~\IEEEmembership{Student Member,~IEEE,}
	Subir Majumder,~\IEEEmembership{Member,~IEEE,}
    Hasan Ibrahim,~\IEEEmembership{Student Member,~IEEE,}
    Prasad Enjeti,~\IEEEmembership{Fellow,~IEEE,}
    Le Xie,~\IEEEmembership{Fellow,~IEEE,}
    Chao Tian~\IEEEmembership{Member,~IEEE}
    \\
    Power Electronics \& Power Quality Laboratory, Dept of ECE, Texas A\&M University, \\ College Station, TX  77845; Corresponding author: enjeti@tamu.edu
}

\maketitle
\begin{abstract}
Foundational Large Language Models (LLMs) such as GPT-3.5-turbo allow users to refine the model based on newer information, known as ``fine-tuning''. This paper leverages this ability to analyze AC-DC converter behaviors, focusing on the ripple current in DC-link capacitors. Capacitors degrade faster under high ripple currents, complicating life monitoring and necessitating preemptive replacements. Using minimal invasive noisy hardware measurements from a full bridge rectifier and 90W Power Factor Correction (PFC) boost converter, an LLM-based models to predict ripple content in DC-link currents was developed which demonstrated the LLMs' ability for near-accurate predictions. This study also highlights data requirements for precise nonlinear power electronic circuit parameter predictions to predict component degradation without any additional sensors. Furthermore, the proposed framework could be extended to any non-linear function mapping problem as well as estimating the capacitor Equivalent Series Resistance (ESR).

\end{abstract}

\begin{IEEEkeywords}
Power electronic Converters, Fine-tuning, Large Language Models (LLMs)
\end{IEEEkeywords}

\section{Introduction}

The 2021 publication \cite{1} effectively summarizes the expanding applications of artificial intelligence (AI), excluding the use of Large Language Models (LLMs), in the field of power electronics. In the recent past, the generative nature of LLMs and their ability to perform various natural-language processing tasks has been garnering significant attention in the scientific and industrial community \cite{2}. LLMs saw application in chip design \cite{3}, in electronic design automation and power converter modulation design\cite{4,5}, and in electric power systems and smart grid \cite{6,7,8,10}. These diverse applications are enabled by transformer models, which are the foundation of LLMs, owing to their efficient and powerful pattern recognition capabilities across various tasks. The influx of investment into LLM research and development highlights the industry's commitment to understanding and leveraging these capabilities. The premise is that if the LLMs are able to predict the next word in a sentence (a token to be specific), they should also be capable of building a predictive model of a nonlinear circuit with reasonable accuracy if the LLM is fine tuned using domain-specific data. Therefore, there is strong potential for LLMs to be used to understand the nonlinear circuit behavior in the field of power electronics, a subject matter of this paper.


\begin{figure}[hbt!]
\centering
    \includegraphics[width=8cm]{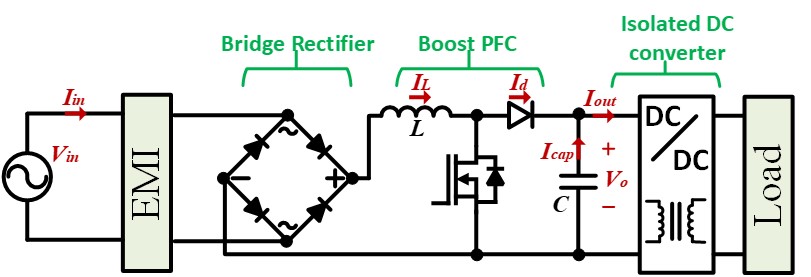}
    \caption{Single phase boost Power Factor Correction (PFC) AC to DC rectifier schematic.} \label{PFC}
    \vspace{-1em}
\end{figure}

Converters based on single-phase Power Factor Correction (PFC) circuits are essential components in a wide range of applications, from laptop chargers to switching power supplies used in large data centers, because they enable unity power factor operations. These converters also have an immense growth potential from \$2.3B in 2023 to \$4.83B by 2031 \cite{360research:PFC}. As depicted in Fig. \ref{PFC}, like other AC/DC or DC/AC converters, DC-link capacitors play an important role in the PFC circuits because they help in smoothing out the voltages as the converters operate in various system conditions. In PFC circuits, DC-link capacitors behave as a decoupling element, bridging the low-frequency 50/60 Hz utility voltage and the high-frequency DC-DC conversion stage. However, these capacitors have a limited life span and can fail prematurely as they are subjected to various stressed operating conditions. One of the common causes of breakdown for power electronic circuits is the failure of capacitors \cite{6357322}. The failure of a capacitor is not always immediately apparent, which can lead to increased strain on the remaining capacitors, accelerating their degradation and leading to eventual failure of the entire converter. 



Capacitor life deteriorates faster under conditions of elevated operating temperatures, high humidity, over-voltage stress, pulsed discharge, and excessive ripple currents. In PFC circuits, capacitors experience heightened ripple currents often exacerbated by real-world operational environments. This increased ripple current results in increased power losses, raising the internal/operating temperature of capacitors. Increased operating temperature increases the Equivalent Series Resistance (ESR) of the capacitors, which gets further increased in a positive feedback loop \cite{5729361}, accelerating the aging process of the DC-link capacitors \cite{5618955}. Hence, understanding and accurately predicting ripple currents (a nonlinear phenomenon) is critical in ensuring the reliability of the power supply units. Previous research has demonstrated the ability to use circuit harmonics in accurate ESR predictions \cite{inproceedings}. 
\begin{figure}[hbt!]
\centering
    \includegraphics[width=7 cm]{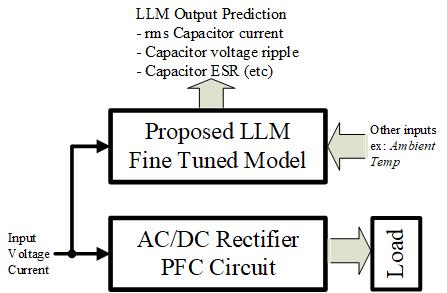}
    \caption{Block diagram of the proposed LLM fine tuned model for the ac/dc rectifier PFC circuit} \label{LLM_Model}
    \vspace{-1em}
\end{figure}
This paper seeks to explore the practical applications of LLMs in predicting key behaviors of power electronic circuits, specifically ripple current content in the DC-link capacitors. Fig. \ref{LLM_Model} shows the block diagram of the proposed prediction approach. Our contributions are therefore twofold:

\begin{itemize}
 \item[(i)] \ul{Use of LLMs for the Prediction of Nonlinear Circuit Behavior:} This study explores the ability of large language models (LLMs) to accurately model nonlinear behaviors in power electronic circuits, specifically bridge rectifier circuits and active PFC boost circuits. This was tested using direct question answering, few-shot prompting, and LLM fine-tuning \cite{7}.

 \item[(ii)] \ul{Data Requirements for Circuit Parameter Prediction:} To develop the models, both simulations and hardware experiments were performed to determine the root mean square (RMS) and the ripple content in the DC-bus capacitor current under various loading conditions. This paper specifically focuses on scenarios where the input waveform is sinusoidal with fixed amplitude and constant ambient temperature. Detailed experiments considering various operating conditions, including the capacitor life estimation, as shown in Fig. \ref{LLM_Model}, will be addressed in future work.
 
\end{itemize}

The remainder of the paper is organized as follows. Section \ref{sec:3} provides a brief overview of the methodology for the estimation of capacitor ripple current, including the experimental setup and measurement techniques. Section \ref{sec:2} first explores the limitations of LLMs in direct question-answering and then introduces the novel concept of `function evaluation' as a prompt prefix to mitigate these limitations. Fine-tuning, alongside in-context learning, has been employed to capture the underlying patterns within the datasets. Section \ref{sec:4} discusses the validation performance of the developed model. Finally, Section \ref{sec:5} concludes this paper.

\section{Capacitor Ripple Current Estimation} \label{sec:3}

\begin{figure}[htb]
     \vspace{-1em}
     \centering
     \includegraphics[width=0.85\linewidth]{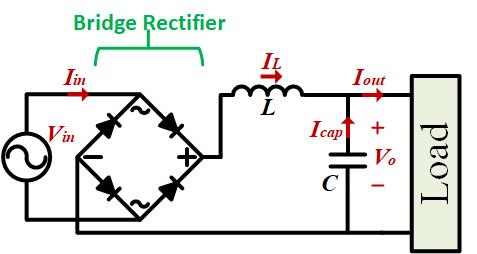}
     \caption{Single phase bridge rectifier schematic, \(V_{in}\), \(V_{o}\), and \(I_{cap}\) are the variables of interest. }
     \label{fig=BR_schenmatic}
     \vspace{-1em}
 \end{figure}

In this study, the capacitor current behavior in two specific power electronic circuits are examined: the bridge rectifier (Fig. \ref{fig=BR_schenmatic}) and the boost PFC circuit (Fig. \ref{PFC}). Through out the experiments, the load is varied for the bridge rectifier while the input voltage is varied for the PFC circuit, fundamentally adjusting the power consumption of both circuits for performing hardware experiments, making power the independent variable. Once the system reaches a steady state, the DC-link capacitor current waveform is captured for both circuits using differential probes. A VARIAC controlled the power supply, and an electronic load is used for the bridge rectifier circuit. Additionally, the bridge rectifier circuit was simulated using PSIM software based on the schematic in Fig. \ref{fig=BR_schenmatic} for preliminary analysis in Section \ref{sec:2}, capturing measurements with virtual probes. After capturing the waveforms for both simulated and hardware circuits, the RMS value is computed and the Fast Fourier Transform (FFT) algorithm is performed in order to obtain the second and fourth harmonics current magnitudes. The detailed test setup will be provided in Section \ref{sec:4}.

\section{Use of LLM ``Fine-tuning'' for Modeling the Behavior of Power Electronic Circuitry} \label{sec:2}

\begin{figure}[htb]
    \centering
    \includegraphics[width=1.0\linewidth]{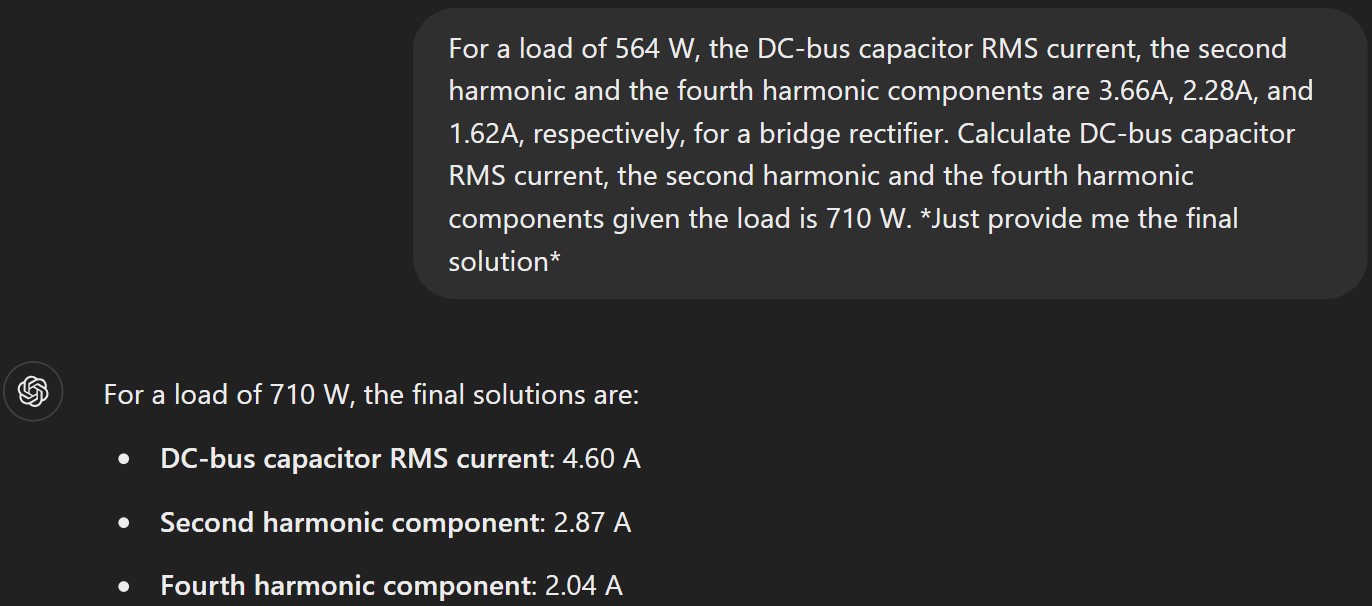}
    \caption{Example of inaccurate output in power electronic circuit analysis when prompted to GPT 3.5. The LLM assumes that the values of interest changes in square root proportion as the system load changes, which is incorrect.}
    \label{fig=Chat_GPT_output}
\end{figure}

The ability to learn from very few labeled samples is a fundamental distinction between machines and humans. LLMs have demonstrated similar capabilities in power engineering applications \cite{7}. However, due to the task-agnostic characteristic of pre-trained LLMs for natural language processing, LLMs perform poorly in application-specific tasks such as circuit analysis even when told about the circuit and provided with sample data as context. This is shown in Fig. \ref{fig=Chat_GPT_output}, where the LLM assumes that the values of interest changes in square root proportion as the system load changes, which is incorrect.


While the specific dataset used for training OpenAI's foundational LLM is unknown, labeling the problem as `circuit theory' consistently hindered the LLMs' ability to recognize patterns within the dataset. To address this, the problem was reframed as an unknown function mapping task. This approach resulted in the LLM generating consistent solutions, enabling it to perform a wide variety of tasks. This, however, also restricts the LLM's capacity to incorporate pre-learned contexts in its output.

\begin{figure}[htb]
    \centering
    \includegraphics[width=0.85\linewidth]{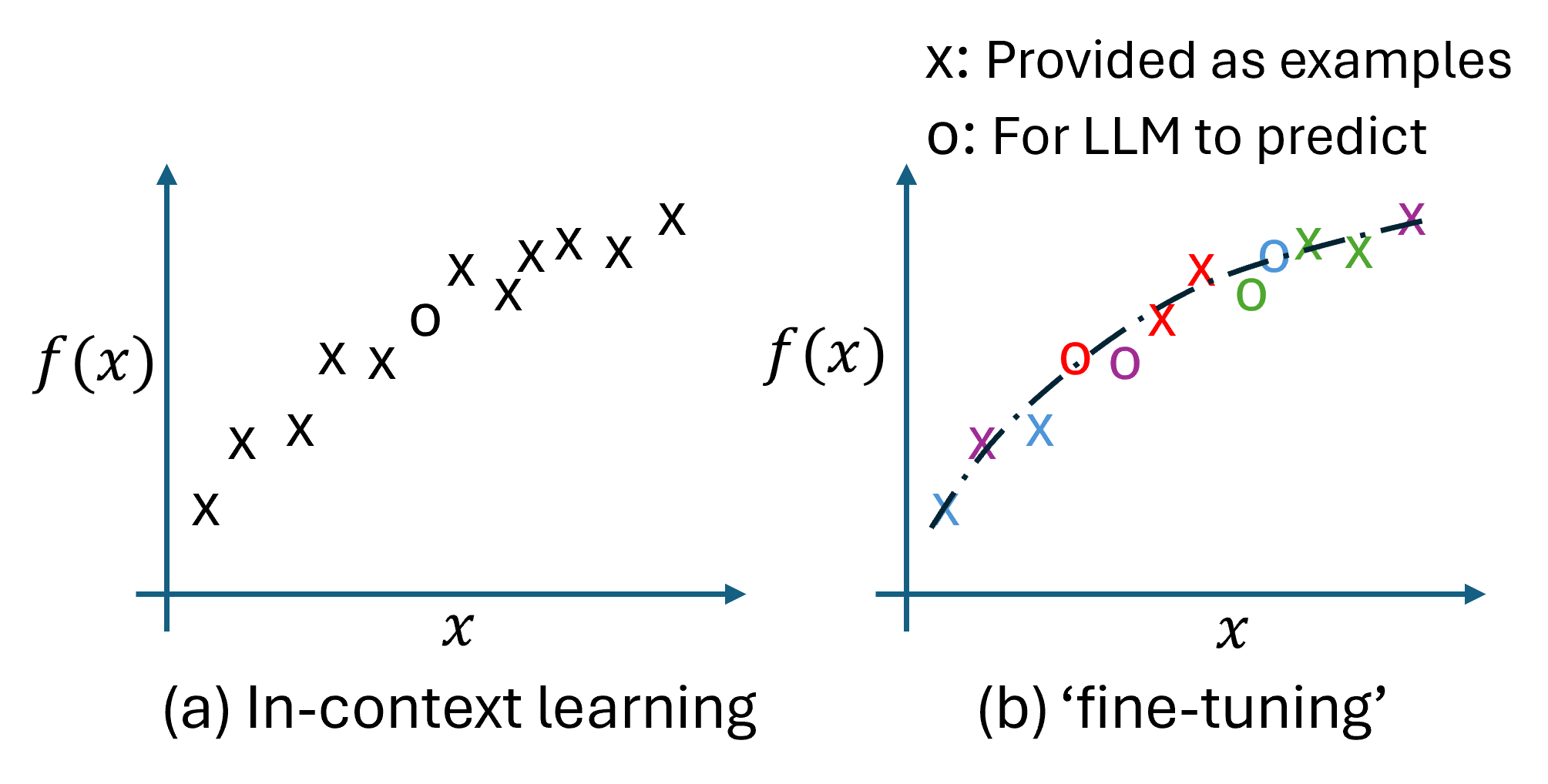}
    \caption{Motivating example for comparing in-context learning and fine-tuning for unknown function mapping.}
    \label{fig=motivation}
\end{figure}

\begin{figure}[htb]
     \centering
     \includegraphics[width=1.00\linewidth]{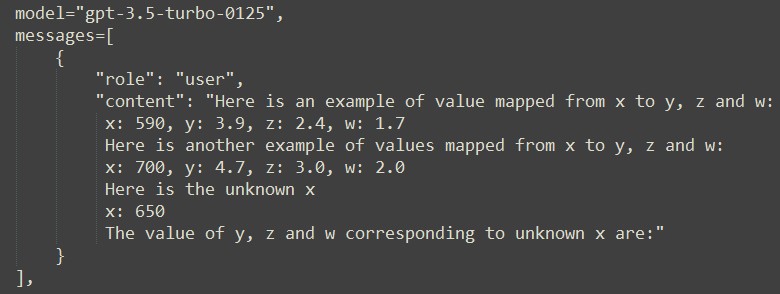}
     \caption{Example of input prompt for the in-context learning.}
     \label{fig=FTexample}
 \end{figure}
\begin{figure}[htb]
    \centering
    \includegraphics[width=0.95\linewidth]{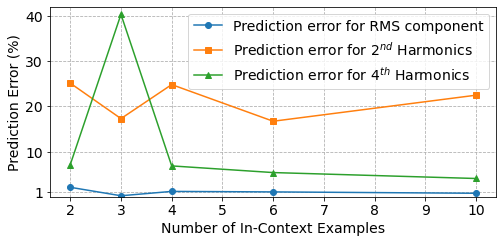}
    \caption{Impact of the number of in-context examples on the prediction accuracy for a given data set.}
    \label{error_examples}
    \vspace{-1em}
\end{figure}

In the context of unknown function mapping, as illustrated in Fig. \ref{fig=motivation}(a), pairs of $x$ and $f(x)$ are provided to the LLM (marked by x) and query it for an unknown value of $x$ (marked by o). In this paper, the aim is to predict capacitor current RMS and harmonic components given different loading conditions, which represents a one-to-many mapping scenario. An example prompt for learning based on in-context examples is demonstrated in Fig. \ref{fig=FTexample}. As shown in Fig. \ref{error_examples}, the accuracy of the LLM's predictions (using GPT-3.5 turbo in this paper) varies non-monotonically with the introduction of more examples. This figure was generated through experimenting with PSIM-generated data, where the input was the load in watts ($x$), and the outputs were the RMS value of capacitor current ($y$) as well as the 2$^{nd}$ ($z$) and 4$^{th}$ harmonic components of currents ($w$) through the DC-bus capacitor. It was observed that the average prediction error (mean absolute percentage error) across multiple experiments for the RMS component and the 4$^{th}$ harmonic component was around 5\% with additional examples. However, the prediction error for the 2$^{nd}$ harmonic component remained high. This example demonstrates the limited ability of the LLMs to learn in-context solely from the examples provided.

The existing literature on LLM provides two major insights. First, fine-tuning outperforms in-context learning without parameter updates \cite{bhatia2023tart}. Fine-tuning involves additional training of pretrained models on specific datasets with labeled data, allowing the model to learn previously unseen patterns and improve its performance on targeted tasks or domains. Second, large language models (LLMs) can be trained to in-context learn a certain function class \cite{garg2022can, bhatia2023tart}. Based on these insights, the following question is posed: 
\uline{"Can LLMs generalize the nonlinear input-output relationship of a system if they are trained using randomly sampled noisy measurements across multiple steps?"}

This process is illustrated in Fig. \ref{fig=motivation}(b). At each step, represented by different colors, the LLM is presented with a few examples and a query point (similar to Fig. \ref{fig=FTexample}), allowing the model to in-context learn from the examples and prior knowledge to predict the function value. The LLM is updated by comparing its estimated value with actual measurements, which provides the LLM with prior knowledge. Random sampling is used across different steps from the training data, meaning the same data points might be shown multiple times (unlike Fig. \ref{fig=motivation}(b), where each measurements are used only once). Based on existing insights, if LLMs can generalize across a function class, they should capture input/output relationships for a given function, even when data points have significant measurement errors. Minibatches, a common technique in machine learning, help achieve this objective. The objective is to generate a model, $M_\theta$, parameterized by $\theta$, such that:

\begin{equation}\label{2}
   \min_\theta \mathbb{E}_P \left[ \frac{1}{k} \sum_{i=1}^{k} \ell_{CE} \left( M_\theta (P_i, x_i^{k+1}), f(x_i^{k+1}) \right) \right]
\end{equation}

Here, at every step, $k+1$ experimental results are randomly sampled, with the first $k$ examples $P_i := (x_1, f(x_1)), \cdots (x_i, f(x_i))$ are chosen as part of the prompt prefix (prompt in Fig. \ref{fig=FTexample} contain three prefixes). The model-generated output for $x_i^{k+1}$ is compared with the true output $f(x_i^{k+1})$, and the difference is captured using the cross-entropy loss function \eqref{2}, $\ell(\cdot)$ \cite{kaplan2020scaling}. The objective for the model is to minimize the expected loss across all prompt prefixes. Given that a production-grade LLM was used, it is difficult to infer which parameters have been updated through the fine-tuning process. However, once the model is trained, its performance is evaluated on the testing set, a subsection of the data which contains unseen examples, in order to assess its performance and ability to generalize.

As in other deep-learning methods, learning by the LLM relies on several hyperparameters, such as the learning rate, batch size, and number of training epochs. Given that the LLMs used here are already foundational, a balanced approach to the fine-tuning is needed. While a small learning rate is important to mitigate the effect of noisy data, it should not degrade the well-tuned weights of the pre-trained LLM. A larger batch size can reduce the variance in the computed gradient; the available hardware resources might also constrain the choice of batch size, making training more expensive. Additionally, in the presence of noisy data, a larger number of training epochs may lead to overfitting the training data, resulting in an increase in the training set performance and a decrease in testing performances. Therefore, one should focus on conservative updates that carefully adapt to new patterns in the data, leading to improved performance on realistic tasks without capitulating to noise-driven inaccuracies.

\begin{figure}[ht]
  \centering
  \begin{subfigure}[b]{0.49\linewidth}
    \centering
    \includegraphics[width=\linewidth]{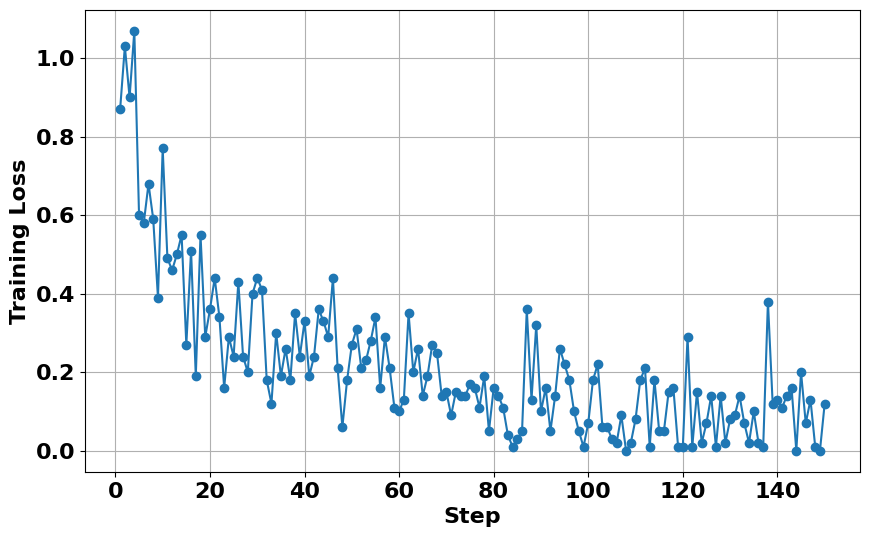}
    \label{BR_E_TL}
    \caption{Full bridge rectifier}
 \end{subfigure}
  \hfill
  \begin{subfigure}[b]{0.49\linewidth}
    \centering
    \includegraphics[width=\linewidth]{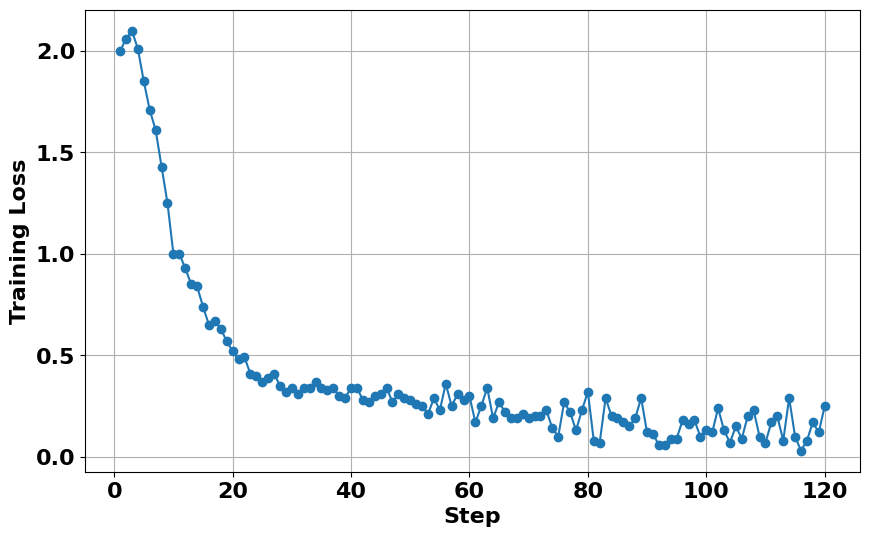}
   \label{BR_S_TL}
    \caption{Active PFC boost circuit}
  \end{subfigure}
  \caption{Training loss \eqref{2} considering experimental data for the full Bridge rectifier and active PFC-boost converter.}
  \label{fig=two_figures}
\end{figure}

Before training with experimental data from both the full-bridge rectifier and PFC boost circuit, the dataset was separated into testing and training set randomly. Hardware measurements from the power electronic circuits were utilized as a part of this exercise. For each prompt, 11 different experimental results were randomly selected from the training set, using 10 as prompt prefixes. The sequential order of examples were assumed to be insignificant, so the total number of possible combinations to show the results to the LLM is $N\mathbb{C}11$, where $N > 11$. Given the large number of combinations, 150 unique combinations were chosen for the training prompts. For the bridge rectifier circuit, a learning rate of 1.5, a batch size of 5, and 5 epochs were used. The same batch size was applied to the PFC circuit, with a learning rate of 1 and 4 epochs. The training loss for models considering hardware experimental data from both circuits is shown in Fig. \ref{fig=two_figures}. Training with both circuits resulted in a significant decrease in training loss within 20 model ($M_\theta$) updates. Although there were fluctuations in the training loss, it is expected that with enough training, the losses could converge closer to zero, which could be explored in future work. It was also observed that for the bridge rectifier model, the losses did not fully converge in 100 iterations, which could negatively impact validation. For the PFC circuit model, higher fluctuations in the training loss indicate the complexity of the training data. However, the eventual convergence of the loss suggests the LLM's ability to generalize unknown functions despite the noisy or complex dataset.

\section{Analysis of AC-DC Rectifier Systems} \label{sec:4}

After training the model, 10 equal spaced points were chosen from the training set as part of the prompt-prefix, and the query points were selected from the testing set for validation. Since LLMs are generative models, their predictions inherently involve some randomness. To address this variability, the model was run 20 times, the mean of which was taken as the model output. All of these generated results are presented in the following subsections, along with the true measurements. The `fine-tuned LLM-generated results were also compared with the results obtained from polynomial regression. The Mean Absolute Percentage Error (MAPE) was used as a metric to calculate the accuracy of both the LLM-based predictions and the regression analysis results, as described in \eqref{1}.

\begin{equation}\label{1}
\text{MAPE} = \left( \frac{1}{n} \sum_{i=1}^n \left| \frac{A_i - F_i}{A_i} \right| \right) \times 100\%
\end{equation}

MAPE measures the average of the absolute differences between predicted and actual values, expressed as a percentage of actual values. Here, \(A_{i} \) represents the actual value, \(F_{i} \) represents the predicted value, and \({n} \) is the number of data points (20 in this case). This metric effectively reflects the magnitude of the error in percentage terms, making it extremely useful for comparing the performance of the model across datasets of different scales.


\begin{figure*}[hbt!]
  \centering
  \begin{subfigure}[b]{0.32\linewidth}
    \centering
    \includegraphics[width=\linewidth]{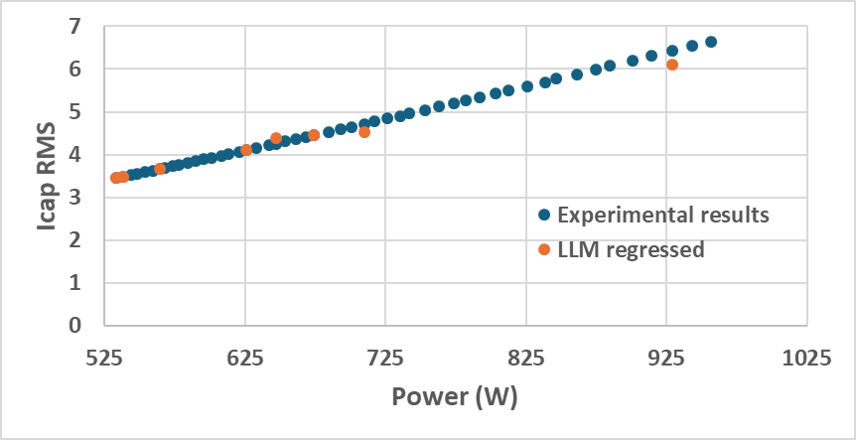}
    \caption{}
    \label{fig=VDC BR}
  \end{subfigure}
  \hfill
  \begin{subfigure}[b]{0.32\linewidth}
    \centering
    \includegraphics[width=\linewidth]{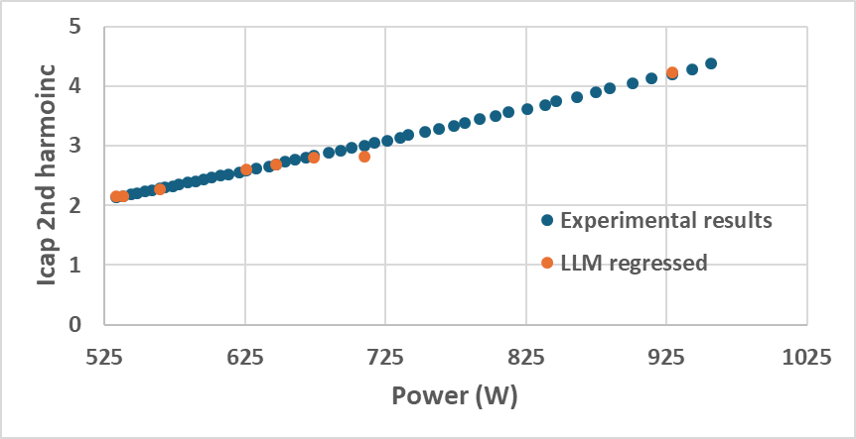}
    \caption{}
\label{fig=second harmonic BR}
  \end{subfigure}
  \hfill
  \begin{subfigure}[b]{0.32\linewidth}
    \centering
    \includegraphics[width=\linewidth]{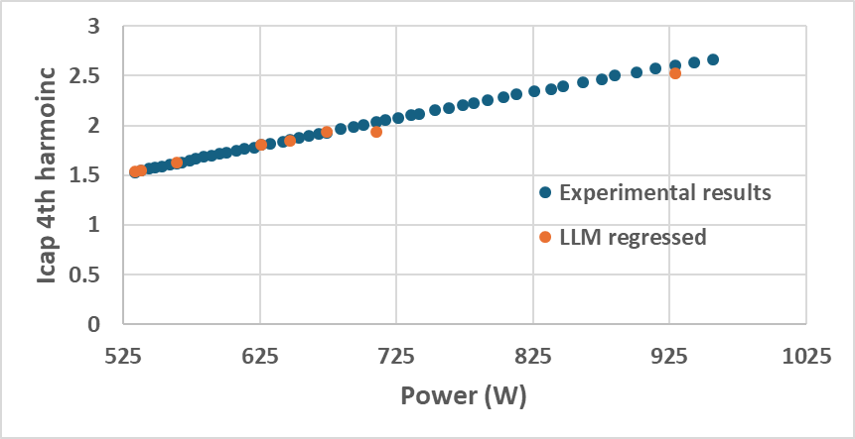}
    \caption{}
    \label{fig=fourth harmonic BR}
  \end{subfigure}
  \caption{Comparing Actual and Predicted values for the experimental FFT fine-tuned predictions for the bridge rectifier (Fig. \ref{fig=BR_schenmatic}). Blue dots
are true values while orange are predicted using the fine-tuning model, all three graphs are smooth and have high prediction accuracy: (a) \(I_{cap}\) RMS values. (b) \(I_{cap}\) second harmonic. (c) \(I_{cap}\) fourth harmonic.}
  \label{fig=BR_results}
\end{figure*}

\subsection{Full-Bridge ac-dc Rectifier Model (Fig.\ref{fig=BR_schenmatic})}

\begin{figure}[htb]
     \centering
     \includegraphics[width=0.99\linewidth]{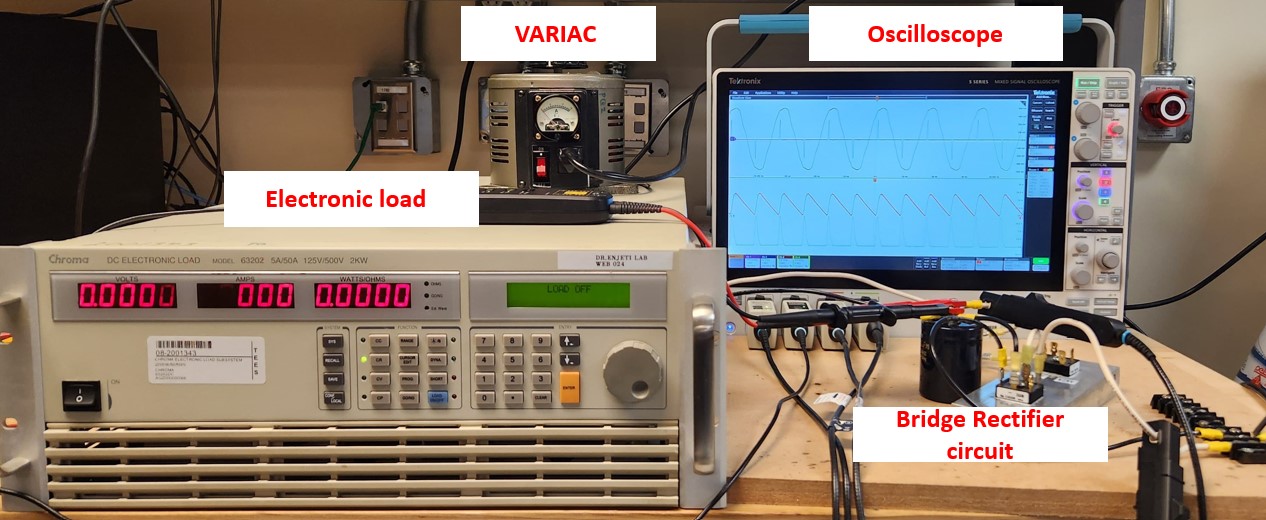}
     \caption{Bridge rectifier (Fig. \ref{fig=BR_schenmatic}) experimental setup. }
     \label{fig=BR_exp_setup}
     \vspace{-1em}

 \end{figure}

The test setup for conducting hardware experiments with the bridge rectifier is illustrated in Fig. \ref{fig=BR_exp_setup}. As detailed in Section \ref{sec:3}, the bridge rectifier circuit is connected to an electronic load for precise load adjustment, a VARIAC for power supply control, and an oscilloscope is used for monitoring voltage and current. Experiments across 50 distinct loading levels were conducted, extracting harmonic components using the Fast Fourier Transform (FFT) algorithm \cite{Nussbaumer1981}. The experimental data is visualized in Fig. \ref{fig=BR_results}. In this figure, the power consumption of the load is used as an independent variable with RMS, 2nd, and 4th harmonic currents as dependent variables. Of all the experimental results, 42 instances were used for training the model, while 8 instances were reserved for testing the model's capabilities.

\begin{table}[htb]
\centering
\caption{Comparing the prediction accuracies of LLM and Linear Regression for Bridge Rectifier Circuit}\label{table=EXP_MAPE1}
\resizebox{\linewidth}{!}{
\begin{tabular}{|c|c|c|c|c|c|c|}  
\hline
Power & \multicolumn{2}{c|}{$I_{cap}$ RMS}  & \multicolumn{2}{c|}{$2^{nd}$ harmonic $I_{cap}$ }  & \multicolumn{2}{c|}{$4^{th}$ harmonic $I_{cap}$} \\
(W) & \multicolumn{2}{c|}{(Absolute Percentage Error}  & \multicolumn{2}{c|}{(Absolute Percentage Error)}  & \multicolumn{2}{c|}{(Absolute Percentage Error} \\ \cline{2-7}
 & LLM & Regression & LLM & Regression & LLM & Regression \\               
\hline
\hline
533& {\color{blue}0.49}& \textbf{0.88} & {\color{blue}0.53}& \textbf{1.44} & {\color{blue}0.64}& \textbf{0.93}\\
\hline
538& {\color{blue}0.19}& 0.86 & {\color{blue}0.07}& 1.16 & {\color{blue}0.05}& 0.81\\
\hline
564& {\color{blue}0.17}& 0.38 & {\color{blue}0.33}& 0.54 & {\color{blue}0.37}& 0.55\\
\hline
626& {\color{blue}0.06}& 0.11 & {\color{blue}0.31}& 0.10 & {\color{blue}0.06}& 0.36\\
\hline
646& 2.82& {\color{blue}0.25} & {\color{blue}0.13}& 0.29& 0.72& {\color{blue}0.23}\\
\hline
674& {\color{blue}0.28}& 0.35& 0.95& {\color{blue}0.31} & 0.58& {\color{blue}0.01}\\
\hline
710& 3.83& {\color{blue}0.56}& \textbf{6.40}& {\color{blue}0.83}& \textbf{4.96}& {\color{blue}0.40}\\
\hline
929& \textbf{4.91} & {\color{blue}0.50} & 0.83& {\color{blue}0.66}& 3.05& {\color{blue}0.60}\\
\hline
\end{tabular}
}
\end{table}

Fig. \ref{fig=BR_results} illustrates the accuracy of the predicted results. As seen in Figure \ref{fig=two_figures}(a), the impact of the training loss, which has not yet converged, is evident in the `fine-tuned' LLM-generated results. Specifically, at power levels of 710 W and 929 W, the LLM performed poorly. This poor performance is also reflected in Table \ref{table=EXP_MAPE1}, where the MAPE increased to 6.4\% for these power levels, compared to below 3\% for the validation dataset between 533 W and 674 W, where the LLMs performed noticeably well. The worst performing validation points for the RMS, as well as the second and fourth harmonic currents for both LLM-based prediction and regression analysis are shown in bold in Table \ref{table=EXP_MAPE1}. While the worst-case MAPE is higher for LLM-based prediction, LLM-based predictions perform well on average (marked in blue color) compared to linear regression analysis. Further investigations are needed to reduce the error across all of the validation set to below 1\%.

\subsection{Single Phase boost PFC AC-DC Converter (Fig. \ref{PFC})}

\begin{figure}[hbt!]
\centering
    \includegraphics[width=8cm]{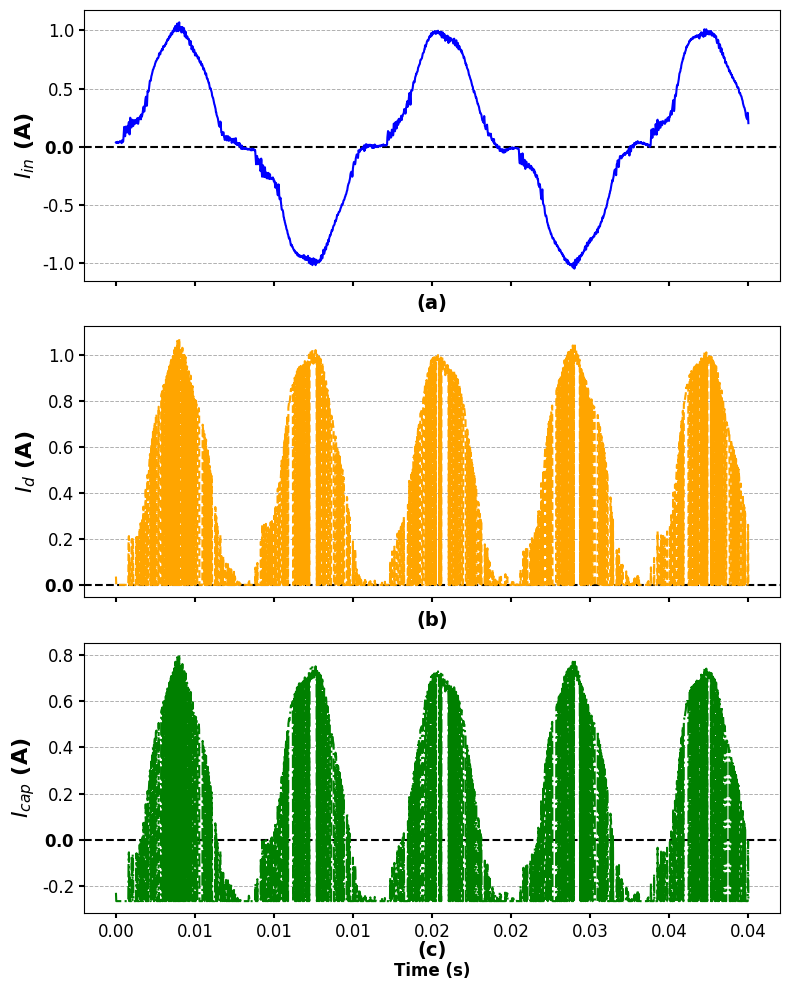}
    \caption{Experimentally measured input current ($I_{in}$) in blue, diode current ($I_{d}$) in orange, and capacitor ripple current ($I_{cap}$) in blue for the single-phase PFC circuit (Fig. \ref{PFC}).} \label{PFC_sample}

\end{figure}

\begin{figure*}[hbt!]
  \centering
  \begin{subfigure}[b]{0.32\linewidth}
    \centering
    \includegraphics[width=\linewidth]{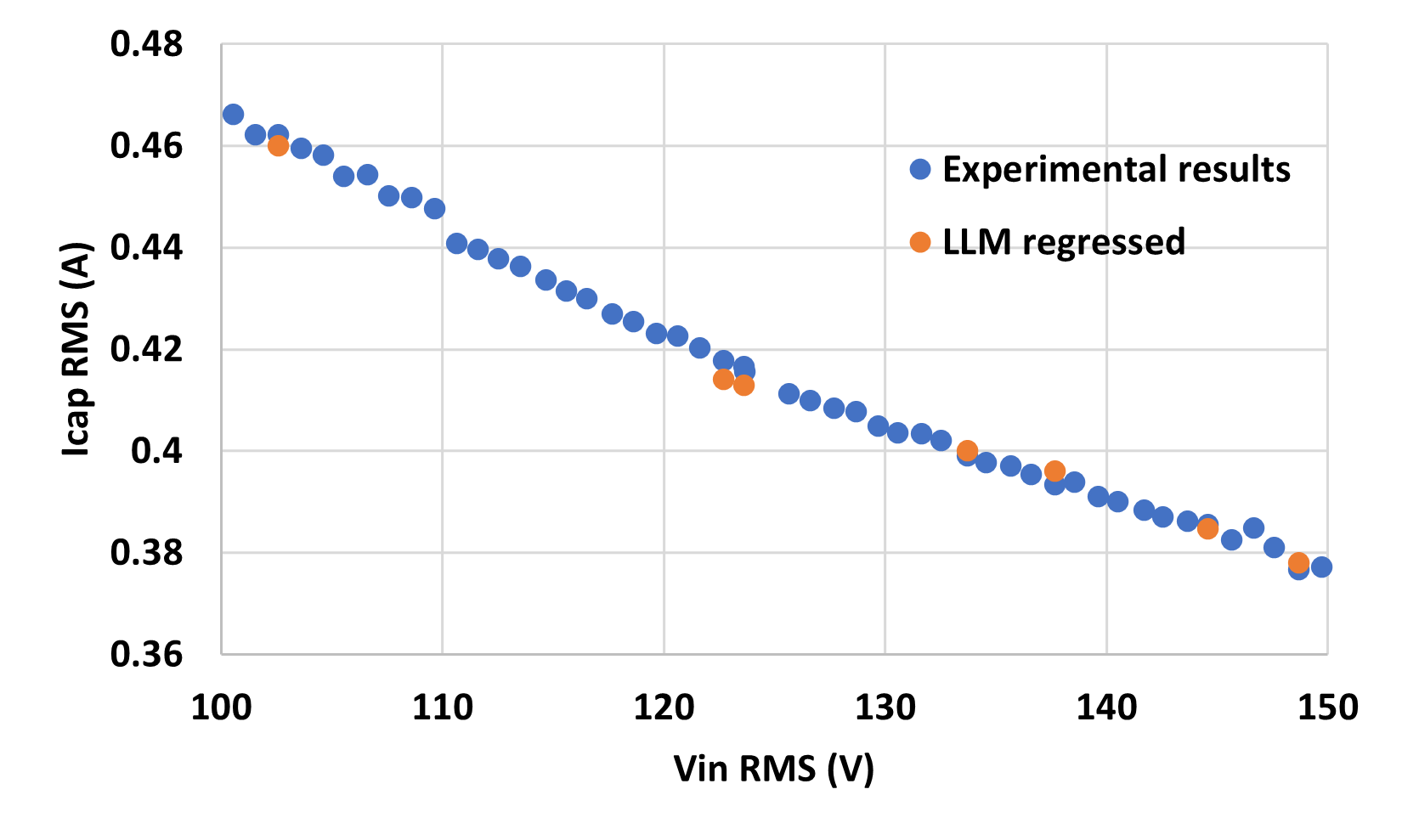}
    \caption{}
    \label{fig=VDC BR}
  \end{subfigure}
  \hfill
  \begin{subfigure}[b]{0.32\linewidth}
    \centering
    \includegraphics[width=\linewidth]{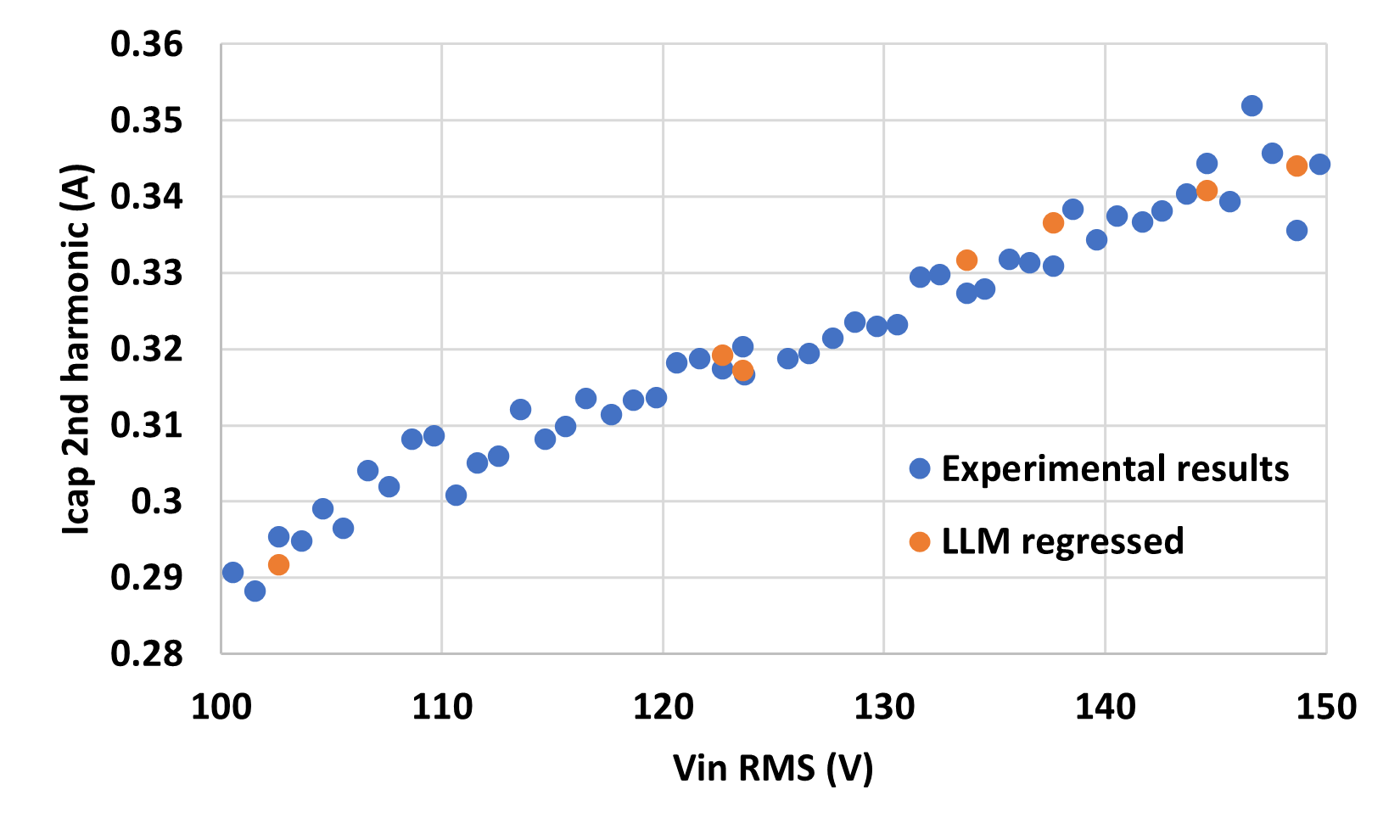}
    \caption{}
\label{fig=second harmonic BR}
  \end{subfigure}
  \hfill
  \begin{subfigure}[b]{0.32\linewidth}
    \centering
    \includegraphics[width=\linewidth]{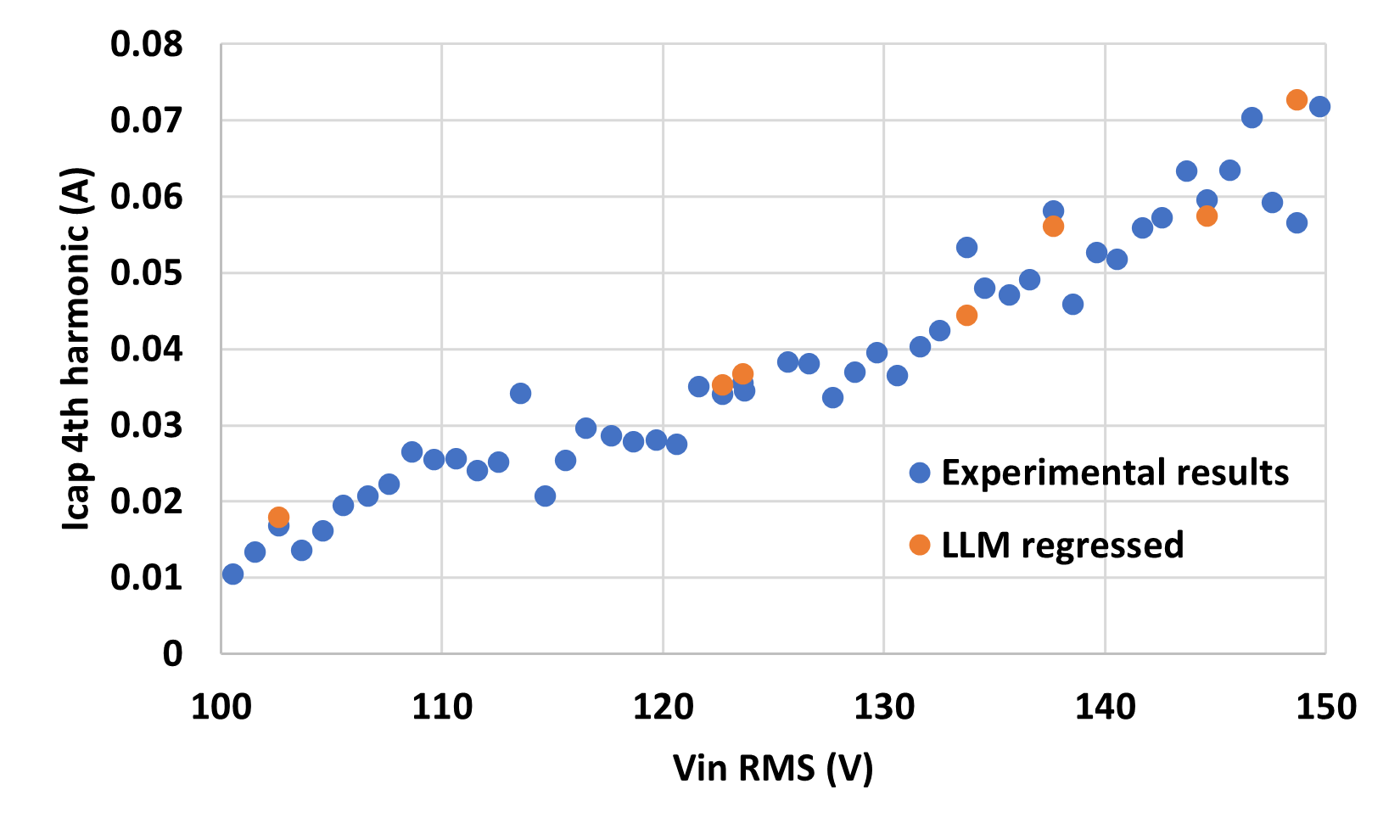}
    \caption{}
    \label{fig=pfc_performance}
  \end{subfigure}
  \caption{Comparing Actual and Predicted values for the experimental FFT fine-tuned predictions for the single phase boost PFC (Fig. \ref{fig=BR_schenmatic}). Blue dots are true values while orange are predicted using the fine-tuning model, all three graphs demonstrate good accuracy despite noise: (a) \(I_{cap}\) RMS values. (b) \(I_{cap}\) second harmonic. (c) \(I_{cap}\) fourth harmonic.}
  \label{fig=PFC_results}
  \vspace{-1em}

\end{figure*}

For the PFC circuit, data was collected while varying the input voltage. The raw data was then processed through FFT to extract the RMS capacitor current, as well as the second and fourth harmonics. Fig. \ref{PFC_sample} presents sample waveforms of the PFC circuit experimental data where the waveform corresponding to the input current ($I_{in}$) is first presented in blue, the diode current ($I_{d}$) in orange, and the current through the capacitor (\(I_{cap}\)) in green . As depicted in the figure, the capacitor current contains significant high-frequency components, which may limit our ability to accurately estimate harmonic components as shown in Fig. \ref{fig=pfc_performance}. In this section, only second and fourth harmonics component prediction was investigated, other high-frequency current measurements could also be included in future work without loss of generality.

\begin{table}[htb]
\centering
\caption{Comparing the prediction accuracies of LLM and XGBoost Regression for PFC Circuit}
\label{table=SIM_PFC_MAPE}
\resizebox{\linewidth}{!}{
\begin{tabular}{|c|c|c|c|c|c|c|}  
\hline
{$V_{in}$} & \multicolumn{2}{c|}{$I_{cap}$ RMS}   & \multicolumn{2}{c|}{$2^{nd}$ harmonic $I_{cap}$ }  & \multicolumn{2}{c|}{$4^{th}$ harmonic $I_{cap}$} \\
(V) & \multicolumn{2}{c|}{(Absolute Percentage Error)}  & \multicolumn{2}{c|}{(Absolute Percentage Error)}  & \multicolumn{2}{c|}{(Absolute Percentage Error)} \\ \cline{2-7}
 & LLM & Regression & LLM & Regression & LLM & Regression \\               
\hline
\hline
102 & {\color{blue}0.45} & 0.71 & {1.25} & \color{blue}0.74 & \color{blue}6.22 & {19.24}  \\
\hline
122 & 0.87 & {\color{blue}0.61} & {0.59} & {\color{blue}{0.16}}  & {3.78} & \color{blue}{2.15}  \\
\hline
123 & 0.91 & {\color{blue}0.86} & {0.99} & \color{blue}0.77 & 3.41 & \color{blue}{2.03} \\
\hline
133 & {\color{blue}0.24} & 1.00  & {1.33} & \color{blue}0.39  & {\color{blue}16.71} & \textbf{20.61}   \\
\hline
137 & {\color{blue}0.66} & {1.00} & {1.71} & {\color{blue}{0.05}} & \color{blue}3.57 & {17.85}  \\
\hline
144 & {\color{blue}0.21} & {0.42} & {\color{blue}1.02} & 1.52  & \color{blue}\textbf{3.57} & {6.07}   \\
\hline
148 & {\color{blue}{0.37}} & 1.82  & \color{blue}\textbf{2.54} & \textbf{2.67}  & \textbf{28.59} & \color{blue}{7.23}   \\
\hline
150 & {\color{blue}\textbf{1.88}} & \textbf{1.92} & \color{blue}0.03 & {0.96} & 19.28 & {\color{blue}6.65}  \\
\hline
\end{tabular}
}
\vspace{-1em}
\end{table}

In Fig. \ref{fig=pfc_performance}, the experimental data is visualized using blue dots, while the LLM-predicted data is shown in orange dots. The XGBoost regressor was used as a benchmark of more traditional ML algorithms. Although both the regression and LLM-based models achieve comparable Mean Absolute Percentage Values for the PFC value predictions for all three variables of interest, the LLM is superior in terms of the RMS prediction and is slightly outperformed in the harmonics predictions as can be noticed by the blue highlighted numbers in Table \ref{table=SIM_PFC_MAPE}, achieveing a MAPE score of 0.69\%, 1.18\%, and 10.64\% for the total RMS, 2nd and 4th harmonics respectively. The high accuracy of the LLM model can likely be attributed to its strong ability to recognize patterns. Exploring the accuracy improvement through additional training, hyper parameter tuning, or prompt engineering is beyond the scope of this paper but will be included in future work. Further investigation will also include model improvement by increasing number of examples compared to the current case which was limited to 150 from $42\mathbb{C}11$ possible choices. Finally, the use of the predicted harmonics for ESR estimation will be extensively explored.

\section{Conclusion} \label{sec:5}
This paper investigates the use of LLMs to predict the behavior of power electronic circuits based on limited measurement data. Notably, the `fine-tuned' LLM models developed for the bridge rectifier and single-phase PFC circuits achieved great accuracy. Specifically: (a) The maximum MAPE across the validation set for the PFC-boost circuitry was less than 2\% for both the RMS and second harmonic data; (b) Experimental datasets obtained using the bridge rectifier circuit model showed an average prediction error of 1.59\% for the RMS current through the capacitor, 1.19\% for the second harmonic, and 1.30\% for the fourth harmonic component of the capacitor current; (c) The LLMs successfully captured the general trend within the dataset; (d) Modeling the prediction task as the mapping of an unknown function suggests broader applicability across different domains, demonstrating function generalization. In addition, the fine tuned model's prompt-based usability is particularly advantageous for simplicity and better general accessibility. In conclusion, the study highlights the potential of using LLMs with training sets to predict \textit{RMS} capacitor ripple currents without the use of invasive sensors. Further details on deducing capacitor ESR and investigation into improving accuracy will be addressed in future work.

\bibliographystyle{IEEEtran}
\bibliography{conference_101719.bib}

\end{document}